\documentclass[useAMS,usenatbib]{mn2e}
\input psfig.tex
\title[Mass-richness scaling]{The scaling relation between
richness and mass of galaxy clusters: a Bayesian approach.}
\author[S. Andreon \& M. A. Hurn]{
S. Andreon,$^1$\thanks{stefano.andreon@brera.inaf.it}, M. A. Hurn,$^2$
\\
$^1$INAF--Osservatorio Astronomico di Brera, Milano, Italy\\
$^2$University of Bath, Department of Mathematical Sciences, Bath, UK\\
}
\date{Accepted ... Received ...}
\pagerange{\pageref{firstpage}--\pageref{lastpage}}
\pubyear{2009}
\begin{document}
\maketitle

\label{firstpage}

\begin{abstract}  We use a sample of 53 galaxy clusters at $0.03 < z < 0.1$
with available masses derived from the caustic technique and with velocity
dispersions computed using 208 galaxies on average per cluster, in order to
investigate the scaling between richness, mass and velocity dispersion.  A
tight scaling between richness and mass is found, with an intrinsic scatter of
only 0.19 dex in mass and with a slope one, i.e. clusters 
which have twice as many galaxies are twice as massive.
When richness is measured without any knowledge of the
cluster mass or linked parameters (such as $r_{200}$), it can predict mass
with an uncertainty of $0.29\pm0.01$ dex. As a mass proxy, richness competes
favourably with both direct measurements of mass given by the caustic method,
which has typically $0.14$ dex errors (vs $0.29$) and X-ray luminosity, which
offers a similar $0.30$ dex uncertainty. The similar performances of X-ray luminosity and richness
in predicting cluster masses has been confirmed using cluster
masses derived from velocity dispersion fixed by numerical 
simulations. These results suggest that cluster masses can be
reliably estimated from simple galaxy counts, at least at the redshift and
masses explored in this work. This has important applications in the
estimation of cosmological parameters from optical cluster surveys, because in
current surveys clusters detected in the optical range outnumber, by at least one
order of magnitude, those detected in X-ray. Our analysis is robust from an
astrophysical perspective because the adopted masses are among the most
hypothesis-parsimonious estimates of cluster mass and from a statistical
perspective, because our Bayesian analysis accounts for terms usually
neglected, such as the Poisson nature of galaxy counts, the 
intrinsic scatter and uncertain errors. The data and 
code used for
the stochastic computation is distributed with the paper.  \end{abstract}


\section{Introduction}

Clusters of galaxies are attracting considerable attention for their cosmological
applications. A conceptually simple observation, such as the number of clusters per
unit volume, is able to put strong constraints on the cosmological parameters
(or their combinations), for example on the equation of state of the dark energy
(e.g. Albrecht et al., 2006, i.e. the Dark Energy Report and references
therein). In essence, both
analytic predictions and gravitational N body simulations give the halo mass
function, $dN/dM/dV$, i.e. the number of halos of mass $M$ per unit halo mass and
universe volume. The number of halos is sensitive to the cosmological parameters
in two ways, linearly (with the cosmic volume) and exponentially (via the growth
function, i.e. how the cluster mass increases with time). Since one can in
principle measure the abundance of the clusters in the Universe, the
comparison of the observed number of clusters to the expected
(cosmologically-dependent) number of halos allows one to constrain the cosmological
parameters. 
This is one of the drivers of many on-going cluster surveys, such
as the South Pole Telescope Survey\footnote{PI Carlstrom,
http://pole.uchicago.edu/} using clusters detected by the Sunayev-Zeldovich
effect, the XMM-Large Scale Survey\footnote{PI Pierre,
http://vela.astro.ulg.ac.be/themes/spatial/xmm/LSS/}, the XMM-Cluster Survey\footnote{PI
Romer, http://xcs-home.org/} using clusters detected by their X-ray
emission, MaxBCG (Koester et al. 2007) and the Red Sequence Cluster
Survey\footnote{PI Yee, http://www.rcs2.org/} using clusters detected by optical
data. More recently, lensing cluster surveys have started (e.g. Berg\'e et al.
2008).

As is known, each experiment measures a combination of cosmological
parameters, rather than the parameters per se. Only the combination of several 
measures from different kinds of experiments is able to break this degeneracy 
in the parameter space, also showing the absence of systematic effects. In this
sense, cluster counting is complementary to other experiments such as the
observations of SNIa, or the measurements of Baryon Acoustic Oscillations and
CMB, etc. This last aspect is very
important in order to test the idea that dark energy is
indeed a new source in Einstein equations rather than
e.g. the manifestation of a different theory of gravity; by
comparing observables which are mainly sensitive to the
growth of structures with tests of the redshift-distance
relation, we can look for inconsistencies that cannot be
explained by dark energy in the form of a new fluid (e.g.
Trotta \& Bower 2006).

The main obstacle to using clusters for cosmological tests is that no
technique is able to yield a direct measure of their masses, but instead they measure
proxies such as the X-ray flux, temperature or Yx (Kravtsov et al. 2006), 
$n_{200}$ (a sort of galaxy richness, see below)
or the Sunayev-Zeldovich decrement. 

The calibration between mass and mass proxy (average relation and
intrinsic scatter) can be achieved either by
specific follow-up observations (more direct, or at least independent,
measures of mass), or by a Bayesian technique called in the astronomical 
context self-calibration (Majumdar \& Mohr, 2004; Gladders et
al. 2007), i.e. basically modelling the relation with generic functions
and marginalising over their parameters.
However, cosmological constraints are much
less tight when determined in the absence of an external measure of
the mass-scaling of the mass proxy. In particular, recent work by Wu
et al. (2008) has emphasised how self-calibration is
hampered by secondary parameters (i.e. the halo formation time and
concentration). Therefore, a direct measurement of the scaling relation
is essential to test the assumption of the self-calibration technique, 
namely to determine the shape of the scatter (currently
Gaussian) and of the scaling (currently linear in log units) and
this is a valuable aim {\em per se.}

The caustic method (Diaferio \& Geller 1997; Diaferio 1999) offers
a robust path to estimating cluster mass.
It relies on the identification in projected phase-space (i.e. in
the plane of line-of-sight velocities and projected cluster-centric
radii, $v,R$) of the envelope defining sharp density contrasts
(i.e. caustics) between the cluster and the field region. The
amplitude of such an envelope is a measure of the mass inside $R$.
Of course, there are other observables available for measuring
cluster masses, but these require additional hypotheses. X-ray-determined
masses require measurements of temperature and surface brightness
profiles and are based on the assumption that the cluster hot gas is
in hydrostatic equilibrium, an assumption that has been questioned in
recent years (e.g. Rasia et al. 2006). Masses derived
using Sunayev-Zeldovich (SZ) decrements additionally assume the
intra-cluster medium is isothermal (e.g. Muchovej et al. 2007).
In this paper, we use caustic masses, i.e. masses derived from the caustic
technique which assumes that galaxies trace the velocity
field. As opposed to the dynamical masses, derived
from the virial theorem (i.e. from the velocity dispersion)
or from the Jeans method, caustic
mass does not require that the cluster is in dynamical
equilibrium (see Rines \& Diaferio 2006 for a discussion). 
 On the other hand, the relative novelty of caustic masses make
them much less studied through numerical
simulations and by comparisons to other mass proxies. For
this reason, we look for systematic errors on caustic
masses and we calibrate the mass-richness scaling with
velocity dispersion and with an additional mass proxy based
on velocity dispersion fixed by numerical simulations.

In this paper we
aim to give the absolute calibration of the relation between
$n200$, the number of red galaxies (brighter
than a specified limit and within a given clustercentric distance) 
and mass. We want also to measure the
scatter of the $n200$ mass proxy and compare 
its performance to the $L_X$ mass proxy.

The mass-richness calibration was partially addressed 
in the pioneering work of maxBCG
(Koester et al. 2007; Rozo et al. 2008 and references 
therein).  Because these works
lack clusters with known masses and $r_{200}$ and their
analysis suffers of circularity ($r_{200}$ is derived for
stack of clusters of a given $n_{200}=n(<r_{200})$, i.e. 
of clusters with a known $r_{200}$),
their calibration is doubtful, and in fact, their
$r_{200}$, used to measure $n200$, is found in
later papers to be on average twice 
as large as the assumed $r_{200}$ radius (e.g. Sheldon et al. 2009;
Becker et al. 2007; Johnston et al. 2007),
i.e. they counted galaxies in a radius too large by a factor of two. 
Furthermore, they found a
redshift dependence when none is assumed to be there by definition
(Rykoff et al. 2008; Becker et al. 2007). 
Our analysis does not share the problems they encountered.

Throughout this paper we assume $\Omega_M=0.3$, $\Omega_\Lambda=0.7$ 
and $H_0=70$ km s$^{-1}$ Mpc$^{-1}$. In this paper, velocity
dispersion, usually denoted as $\sigma_v$ in the literature, is
denoted with the symbol $s$. All quantities are measured
in the usual units: velocity dispersions in km s$^{-1}$, 
cluster radii in kpc, X-ray luminosities in erg s$^{-1}$, 
cluster masses
in solar mass units.

\section{Parameter estimation in Bayesian Inference}

The Bayesian approach to statistics has become increasingly popular
over the past few decades as computational and algorithmic advances have 
permitted the analysis of more complex
data sets and the use of more flexible models.
For the theoretician, there are interesting philosophical differences
to be explored between the Bayesian and frequentist approaches.
For the practictioner, Bayesian data analysis provides an additional
valuable statistics tool.
A good introduction to the Bayesian framework can be found in many textbooks
(e.g. Mackay 2003, D'Agostini 2003 and Gelman et al. 2003). 
In this section we will summarise a Bayesian approach to an applied problem.

Suppose one is interested in estimating the (log) mass of a galaxy cluster,
$lgM$.
In advance of collecting any data, we may have certain 
beliefs and expectations about the values of $lgM$.
In fact, these thoughts are often used in deciding which 
instrument will be used to gather data and how this
instrument may be configured. For example, if we are wanting to
measure the mass of a poor cluster via the virial theorem, a Jeans
analysis or the caustic technique, we will select a spectroscopic
set up with adequate resolution, in order to avoid that velocity
errors are comparable to, or larger than, the likely low
velocity dispersion of poor clusters.
Crystalising these thoughts in the form of a probability distribution for
$lgM$ provides the prior $p(lgM)$, used, as mentioned,
in the feasibility section of the telescope time proposal, where 
instrument, configuration and exposure time are set.

For example one may believe (e.g. from the cluster being somewhat poor)
that the log of the cluster mass is probably not far from $13$, plus or minus 1;
this might be modelled by saying that the probability distribution of the
log mass, here denoted $lgM$ is a Gaussian centred on $13$ and with $\sigma$, the 
standard deviation, equal to $0.5$, i.e. $lgM \sim \mathcal{N} (13,0.5^2)$.

Once the appropriate instrument and its set up have been selected, 
data can be collected on the quantities of interest.
In our example, this means we record a measurement of log mass, say
$obslgM200$, via, for example, a caustic analysis, i.e. measuring
distances and velocities. The physics or, sometimes simulations, 
of the measuring process may allow us to estimate the reliability
of such measurements. Repeated 
measurements are also extremely useful for assessing it.
The likelihood is the model which we adopt for how the noisy observation 
$obslgM200$ arises given a value of $lgM$.
For example, we may find that the measurement technique allows us to measure
masses in an unbiased way but with a standard error of 0.1 and
that the error structure is Gaussian, ie.
$obslgM200 \sim \mathcal{N} (lgM,0.1^2)$.
If we observe $obslgM200=13.3$ we usually summarise the above by writing
$lgM=13.3\pm 0.1$.

How do we update our beliefs about the unobserved log mass $lgM$ in light
of the observed measurement, $obslgM200$? 
Expressing this probabilistically, what is the posterior distribution
of $lgM$ given $obslgM200$, i.e. $p(lgM \ | \ obslgM200)$?
Bayes Theorem (Bayes 1764 and Laplace 1812) tells us that
\begin{equation}
p(lgM \ | \ obslgM200) = \frac{ p(obslgM200 \ |\ lgM) p(lgM)} {p(obslgM200)}
\end{equation}
The denominator $p(obslgM200)$, known as the
(Bayesian) evidence, is equal to the integral of the numerator
\begin{equation}
p(obslgM200) = \int p(obslgM200 \ |\ lgM) p(lgM) dlgM
\end{equation}
Notice that, as with frequentist statistical approaches, assumptions have 
been made which should be assessed; neither priors nor likelihoods (on which
frequentist methods such as maximum likelihood estimation is based) are
set in stone.

Simple algebra shows, that in our example the posterior distribution of
$lgM \ | \ obslgM200$ is
Gaussian, with mean
$\mu=\frac{13.0/0.5^2+13.3/0.1^2}{1/0.5^2+1/0.1^2}=13.29$
and $\sigma^2=\frac{1}{1/0.5^2+1/0.1^2}=0.0096$.
$\mu$ is just the usual weighted average of
two ``input" values, the prior and the observation, with weights
given by prior and observation $\sigma$'s. 

In our example, the posterior mean and standard deviation
are numerically almost indistinguishable from the observed
value and its quoted error, however, this is not the rule for
complex data analysis, for example when biases are there 
or in frontier measurements, like in 
Butcher-Oemler studies, where one often finds observed values outside the
range of acceptable values (see, e.g. Andreon et al. 2006). 
From a computational point of view, only simple examples
such as the one described above can generally be tackled analytically.
Markov Chain Monte Carlo (MCMC) methods are widely used for more complex
problems. 

Although this might sound intimidating to the astronomical end-user,
the advent of BUGS-like programs (Spiegelhalter et al. 1996) such as
JAGS (Plummer 2008), allow scientists to apply these ideas
for quite complicated models using a simple syntax. 
In our example, we just need to write in an ascii file the symbolic
expression of the prior, $lgM \sim \mathcal{N} (13,0.5^2)$ and
likelihood, $obslgM200 \sim \mathcal{N} (lgM,0.1^2)$ and nothing more.
For the work in this paper, the JAGS code 
is given in the appendix.

\begin{figure*}
\psfig{figure=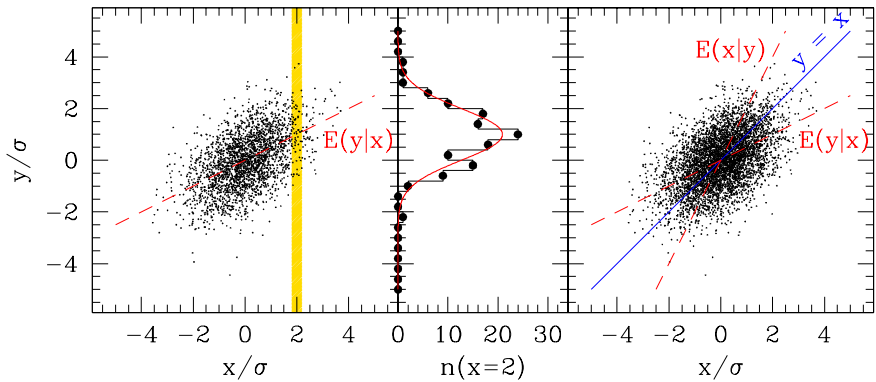,width=12truecm,clip=}
\caption[h]{Left panel: 500 points drawn from a bivariate Gaussian,
overlaid by the line showing the expected value of $y$ given $x$.
The yellow vertical stripe captures those $y$ for which $x$ is
close to 2.
Central panel: Distribution of the $y$ values for $x$ values in a narrow band of $x$ centred on $2$, as shaded in the left panel. 
Right panel: as the left panel, but we also add the lines joining the
expected $x$ values at a given $y$, and the $x=y$ line.
}
\end{figure*}

\section{Uncertainties of predicted values in Bayesian Inference}

Suppose we want to estimate the value of a quantity not yet measured (e.g.
the mass of a not yet weighted cluster). 
Before data $lgM$ are collected (or even considered), the distribution of the 
predicted values $\widetilde{lgM}$ can be expressed

\begin{equation} 
p( \widetilde{lgM} ) = \int p( \widetilde{lgM}, \theta ) d \theta = \int  p( \widetilde{lgM} | \theta ) p(\theta) d \theta
\end{equation} 

These two equalities result from the application of probability
definitions, the first equality is simply that a marginal distribution results
from integrating over a joint distribution,
the second one is Bayes' rule. 

If some data $lgM$ have been already collected for similar objects, we can
use these data to improve our prediction for $\widetilde{lgM}$.
For example, if mass and richness in clusters are highly correlated, 
one may better predict
the cluster mass knowing its richness than in the absence of such 
information, simply because
mass shows a lower scatter at a given richness than when
clusters of all richnesses are considered (except if the relationship
has slope exactly equal to $\tan k \pi/2$, with $k=0,1,2,3$). 
In making explicit the presence of such data, $lgM$,
we rewrite Eq. 3  conditioning on $lgM$:

\begin{equation} 
p( \widetilde{lgM} | lgM ) = \int p( \widetilde{lgM} | lgM, \theta ) p(\theta|lgM) d \theta 
\end{equation} 

The conditioning on $lgM$ in the first term in the integral simplifies out 
because $lgM$ and $\widetilde{lgM}$ are considered conditionally independent given
$\theta$, so that this term becomes simply $p( \widetilde{lgM} | \theta )$. The left
hand side of the equation is called the posterior predictive distribution 
for a
new unobserved $\widetilde{lgM}$ given observed data $lgM$ and model parameters
$\theta$.  Its width is a measure of the uncertainty of the predicted value
$\widetilde{lgM}$, a narrower distribution indicating a more precise prediction. 

Let us first consider a simple example. Suppose we do not know
the mass, $\widetilde{lgM}$, of a given cluster and we
are interested in predicting it from our knowledge of its richness.
In this didactical example we assume for simplicity that a) all 
probability distributions are Gaussian,
b) that previous data $lgM$ for clusters of the same richness
allowed us to determine that clusters of that richness have on average
a mass of $lgM=13.3\pm0.1$,  
i.e. $p(\theta|lgM) = \mathcal{N}(13.3,0.1^2)$,
c) that the scatter between the individual 
and the average mass of the clusters 
is $0.5$ dex, i.e. $p( \widetilde{lgM} | \theta ) = \mathcal{N}(\theta,0.5^2)$. Then,
Eq. 4 is easily analytically solvable and gives 
the intuitive solution that
$p( \widetilde{lgM} | lgM )$ is a Gaussian centred on $lgM=13.3$ and with
a $\sigma$ given by the sum in quadrature of
$0.1$ and $0.5$ ($=0.51$ dex). Therefore, a not-yet
weighed cluster of the considered richness has
a predicted mass of $13.3$ with an uncertainty of $0.51$ dex.
The latter is the
performance of richness as a mass estimator in our didactical
example. A different proxy, say X-ray luminosity, may give a different
value for the uncertainty of the predicted mass and the comparison
of these values allows us to rank the performances of these different
mass proxies.

Later in this paper, we measure and compare the performance
of mass and X-ray luminosity.
The assumptions we use then go beyond the simplistic ones of
the pedagogical example, 
starting with the assumption of
having a set of clusters with richness identical to that
of the cluster whose mass we want to estimate, the
(tacit) assumption of living in an observational error-free world, the lack
of modelling of a trend between richness and mass, the perfect
knowledge of the parameters of the sampling distribution, a perfect
matching of the richness of clusters with available mass and those
with to-be-estimated mass, etc. 
Despite this apparent complexity, to account for all these factors,
we only need to state a richness-mass scaling model (the
same one used to analyse the scaling itself, detailed in Sec 5.1) and use
Eq. 4 to measure the performances of the mass proxies.

Although the above methodology might appear initially intimidating to the 
astronomical
end-user, the use of predictive posterior distributions is generally pain free
since programs such as BUGS offer it as a standard feature. 
In practice,
the integral in Eq. 4 is computed
quite simply using sampling; repeatedly values of $\theta$ are drawn
from the posterior $p(\theta|lgM)$ and for each of these, values of
$\widetilde{lgM}$ are drawn from $p(\widetilde{lgM}|\theta)$. The values of
$\widetilde{lgM}$ are stored. The width of
the distribution of these values gives the uncertainty of the predicted value,
i.e. the performance of the considered mass proxy. Therefore,
the quoted performance accounts for
all terms entering into the modelling of proxy and mass, 
which include the uncertainty of the proxy value
(richness and X-ray luminosity), the uncertainty on the parameters
describing the regression between mass and mass proxy (slope, intercept, 
intrinsic scatter and their covariance), as well as other modelled terms (we
also account for the noisiness of the error itself in our analysis).
Some factors are automatically accounted for without any additional input,
for example, where data are scarce, for example near or outside the sampled
richness or $L_X$ range, predictions are noisier (because the
regression is poorly determined here). As a consequence, proxy performances
are poorer (the posterior predictive distribution is wider) there.

\section{Prediction with errors on predictor variables}

It is important to distinguish between the prediction of a 
variable $y$ which is assumed to be linearly related to
a {\bf non-random} predictor variable $x$ and the prediction
of a variable $y$ which is linearly related to a predictor
variable $x$ which is itself a random variable. 
The latter situation is the one in which we find ourselves here,
given that we want to predict mass as a function of richness
and for both quantities we must collect observational
data.

Figure 1 shows a set of 500 points drawn from a
bivariate Gaussian where marginally both $x$ and $y$ are
standard Gaussian with mean 0 and variance 1 and $x$ and $y$
have correlation $1/2$.
Superimposed on the left hand panel of Figure 1 is the
line giving the theoretical conditional expectation of
$y$ given $x$ (this is known theoretically for this bivariate
Gaussian to be $y = 0.5 x$).
By eye, this line perhaps seems too shallow with respect to 
the trend identified by the points, which perhaps might be 
captured by the $x=y$ line shown in blue in the right-hand panel.
However, if what we want to do is to predict a $y$ given an $x$
value, this ``too shallow'' line is more appropriate.
To illustrate why this is the case, the middle panel of Figure 1
concentrates on those observed for $x$ close to 2.
It is clear from their histogram that their average is closer
to the value predicted by the red line (1 in this case) than the value predicted by the blue (2 in this case).
To emphasise that although we treat $x$ and $y$ symmetrically
in terms of both being random variables, we have an asymmetry
in terms of our predictive goals, the right hand panel also 
shows the expected value of $x$ given a value of $y$.

Akritas \& Bershady (1996) give a related description of
the various types of fit from a non-Bayesian perspective.

\section{Data \& data reduction}


\begin{table*}
\caption{Observed galaxy counts and solid angle ratios.
Column two and three lists the observed galaxy counts in the cluster 
and control directions. The latter subtend a solid angle $C$ times larger
than the former. 
Columns five and six repeat the content of column
two and four, but for a different cluster solid angle, whose radius is
determined by eq. 16, which uses the galaxy counts listed in column 7.}
\begin{tabular}{l r r r r r r}
\hline
Cluster id &   $obstot$ & $obsbkg$ & $C$ & $\widehat{obstot}$ & $\widehat{C}$ & $obsn(<1.43)$ \\ 
(1) & (2) & (3) & (4) & (5) & (6) & (7) \\
\hline
A0160              &  28  &  13   & 3.107 &  29 &   2.951 &   31  \\      
A0602              &  45  &  37   & 3.186 &  23 &   10.77 &   29  \\      
A0671              &  44  &  20   & 2.545 &  36 &   5.443 &   37  \\      
A0779              &  27  &   0   &0.2245 &  19 &  0.4303 &   29  \\      
A0957              &  33  &  20   & 4.605 &  26 &   7.947 &   29  \\      
A0954              &  27  & 168   & 44.87 &  28 &   41.05 &   29  \\      
A0971              &  63  & 127   & 9.957 &  50 &   19.57 &   47  \\      
RXCJ1022.0+3830    &  30  &  28   & 8.598 &  26 &   10.27 &   33  \\      
A1066              &  76  &  41   & 4.075 &  65 &   6.421 &   58  \\      
RXJ1053.7+5450     &  48  &  70   & 6.381 &  40 &   13.77 &   38  \\      
A1142              &  25  &   1   & 0.596 &  15 &   1.606 &   24  \\      
A1173              &  28  & 110   & 20.96 &  27 &   30.45 &   25  \\      
A1190              &  65  &  88   & 9.149 &  63 &   9.896 &   55  \\      
A1205              &  46  &  67   & 9.111 &  42 &   11.84 &   43  \\      
RXCJ1115.5+5426    &  52  &  50   & 6.798 &  45 &   10.48 &   43  \\      
SHK352             &  44  &  24   & 3.063 &  32 &   6.125 &   35  \\      
A1314              &  37  &   5   & 1.151 &  33 &   1.832 &   33  \\      
A1377              &  47  &  48   & 6.697 &  50 &   5.913 &   47  \\      
A1424              &  49  &  45   & 8.575 &  39 &   13.47 &   43  \\      
A1436              &  46  &  51   &  15.6 &  64 &   8.021 &   58  \\      
MKW4               &  26  &   1   &0.1811 &  19 &  0.5456 &   19  \\      
RXCJ1210.3+0523    &  30  &  67   & 25.68 &  36 &   19.22 &   38  \\      
Zw1215.1+0400      &  82  &  62   & 10.46 &  90 &   7.965 &   74  \\      
A1552              &  70  & 113   & 15.72 &  78 &   11.33 &   66  \\      
A1663              &  68  &  86   & 7.363 &  55 &   12.23 &   51  \\      
MS1306             &  22  & 104   & 34.78 &  19 &   44.83 &   21  \\      
A1728              &  46  & 135   & 11.93 &  22 &    26.7 &   33  \\      
RXJ1326.2+0013     &  16  & 118   & 34.21 &  12 &   57.11 &   17  \\      
MKW11              &  13  &   8   & 1.927 &   9 &   4.284 &   10  \\      
A1750              &  58  &  86   & 16.57 &  71 &   12.32 &    59  \\      
A1767              &  90  &  35   & 3.314 &  59 &   6.624 &   54  \\      
A1773              &  52  &  90   & 12.51 &  49 &   15.08 &   45  \\      
RXCJ1351.7+4622    &  18  &  31   & 25.54 &  29 &   13.96 &   35  \\      
A1809              &  63  & 121   & 16.62 &  67 &   11.66 &   56  \\      
A1885              &  29  &  74   & 9.011 &  21 &   50.58 &   20  \\      
MKW8               &  19  &   8   & 2.823 &  17 &    3.39 &   19  \\      
A2064              &  30  &  47   & 11.97 &  22 &   21.44 &   29  \\      
A2061              &  95  &  80   & 5.412 &  85 &   7.381 &   69  \\      
A2067              &  24  & 128   & 37.06 &  28 &    24.3 &   31  \\      
A2110              &  39  & 176   & 21.44 &  32 &   34.32 &   33  \\      
A2124              &  70  &  29   & 2.492 &  48 &   6.036 &   47  \\      
A2142              & 141  & 115   & 10.83 & 186 &   6.141 &   113  \\      
NGC6107            &  28  &  10   & 2.195 &  22 &   4.034 &   22  \\      
A2175              &  49  &  77   & 35.08 &  71 &   14.64 &   66 \\      
A2197              &  35  &   3   & 1.814 &  63 &  0.8029 &   59  \\      
A2199              &  77  &   0   &0.3236 &  88 &   0.239 &   75  \\      
A2245              &  94  &  80   & 6.411 &  88 &   8.376 &   73  \\      
A2244              &  99  & 112   &  11.9 &  99 &   11.75 &   82  \\      
A2255              & 167  &  60   & 3.933 & 173 &   3.514 &  121  \\      
NGC6338            &  26  &   2   &0.3734 &  16 &   1.068 &   19  \\      
A2399              &  47  &  48   & 10.82 &  56 &   7.135 &   51  \\      
A2428              &  37  & 154   & 18.16 &  33 &    25.2 &   33  \\      
A2670              &  95  &  41   & 9.163 & 109 &   4.442 &   93  \\      
\hline         
\end{tabular} 
\end{table*}

Our starting point is the CIRS (Cluster Infall 
Regions in SDSS, Rines \& Diaferio 2006) cluster catalogue.
Fundamentally, clusters are: a) X-ray flux-selected 
b) with an upper cut at redshift $z=0.1$ (to allow a good caustic
measurement) and c) are in the SDSS DR6 spectroscopic survey.
These catalogues give cluster centres, 
virial radii $r_{200}$ and masses within $r_{200}$,
$M_{200}$, derived by the caustic technique.
CIRS also lists
the cluster velocity dispersion, computed using just those
galaxies inside the caustic, and the turnaround radius.
The velocity dispersions are
computed using, on average, 208 member galaxies per cluster.
We note that in CIRS velocity dispersions are quoted 
with slightly asymmetric errors.
D'Agostini (2004) suggests adopting the average 
of the asymmetric errors as a point value of the error and the
midpoint between the upper and lower values as a point value of the
measurement (velocity dispersion) itself. Masses, as quoted by 
CIRS have more asymmetric errors and are
such that the lower error bar includes negative mass for
some clusters. 
This is compatible with symmetric errors on a log scale being
transformed onto a linear scale and is
supported by the way in which Rines \& Diaferio (2008) 
summarise in their introduction their previous (CIRS) 
paper. Therefore, we convert errors back on the log scale. 
Our statistical analysis accounts for noisiness 
of mass and velocity dispersion estimated errors.

For each cluster, we extract the galaxy catalogues 
from the Sloan Digital Sky Survey (hereafter
SDSS) $6^{th}$ data release (Adelman-McCarthy et al., 2008),
discarding both clusters at $z<0.03$ to avoid shredding problems (large
galaxies are split in many smaller sources)
and two cluster pairs (requiring a deblending algorithm for
estimating the richness of each cluster component).
We also discard clusters not 
wholly enclosed inside the SDSS footprint and a few clusters with hierarchical
centres that have converged on a secondary galaxy clump, instead
of on the main cluster.
One further cluster, the NGC4325 group, 
has been removed because it is of very low richness (it has only two galaxies
brighter than the adopted luminosity limit), far lower than the
other clusters in the sample.
The list of the 53 remaining clusters is given in Table 1.
We emphasise that only two cluster pairs
have been removed from the original sample because of their
morphology, all the other excluded clusters 
have been removed because they are not fully enclosed
in the sky area observed by SDSS or have suspect masses because
the CIRS algorithm converged on a secondary clump.

Basically, 
we want to count red members within a specified luminosity range and
colour and within a given cluster-centric radius, typically $r_{200}$,
as is already done for other clusters at similar redshift
(e.g. Andreon et al. 2006) or in the distant universe (Andreon 2006; 2008;
Andreon et al. 2008b).
We only consider red galaxies because these objects
are those whose luminosity evolution is better known and because
their star formation rate (and therefore luminosity) cannot be altered 
by cluster merging, these objects having already exhausted the barionic 
reservoir needed to form new stars.

Since we aim to replicate the present analysis to include 
additional clusters in future papers, we 
take a (passive evolving) limiting magnitude of $M_V=-20$ mag, which is
the approximate completeness of the SDSS at $z=0.3$ and of the
CFHTLS wide survey and CTIO imaging (e.g. Andreon et al. 2004a) of
the XMM-LSS field at $z\sim1$; it is also a widely
used magnitude cut (e.g. de Lucia et al. 2007, Andreon 2008, etc.).
Magnitudes are passively evolving, modelled with a simple
stellar population of solar metallicity, Salpeter IMF, from Bruzual \&
Charlot (2003), as in De Lucia et al. (2007), Andreon (2008) amongst others.
Such a correction is applied for consistency
with other (past and future) work, but is actually unnecessary for our clusters
because it is negligible given the small redshift range ($0.03<z<0.1$) probed 
in this work.

We count only red galaxies, defined as those 
within 0.1 redward and 0.2 blueward in $g-r$ of the colour--magnitude relation. 
This definition of ``red" is quite simple because for our cluster
sample the resulting number hardly
depends on the details of the ``red" definition; the 
determination of the precise location of the colour--magnitude relation 
is irrelevant because the latter is much 
narrower than 0.3 mag and because practically all 
galaxies brighter than the adopted luminosity cut are red. 
Colours are corrected for the colour--magnitude 
slope, but again this is a negligible correction given the small 
magnitude range explored (a couple of magnitudes).
For the colour centre, we took the peak of the colour distribution. 

Some of the galaxies counted in the cluster line of sight,
are  actually in the cluster fore/background.
The contribution from background galaxies is estimated, as usual, from
a reference direction (e.g. Zwicky 1957; Oemler 1974; Andreon, Punzi \& Grado
2005). The reference direction
is taken outside the turnaround radius, or for the few
clusters too close or near to a SDSS border, near the turnaround radius.

Since richness is based on galaxy counts, it is computed
within a cylinder of radius $r_{200}$. Masses are instead calculated (by
Rines \& Diaferio 2006) within spheres of radius $r_{200}$. 

Table 1 gives for our 53 clusters: 1) the cluster id; 2) the observed number of
galaxies in the cluster line of sight within $r_{200}$, $obstot_i$; 3) the
observed number of galaxies in the reference line of sight, $obsbkg_i$; 4) the
ratio between the cluster and reference solid angles, $C_i$. Columns 5 and 6 list
$obstot_i$ and $C_i$, but for the radius inferred using eq. 18, introduced in sec
5.1, based on the observed number of galaxies, within an aperture of 1 $h^{-1}$
Mpc,  $obsn(<1.43)$. Column 7 lists $obsn(<1.43)$.

\section{Results}

\subsection{Richness-mass model}

The aim of this section is to present a Bayesian analysis of the
richness-mass model.
In particular, we wish to acknowledge the uncertainty in all the
measurements, including in error estimates. Most previous approaches
assume that errors are perfectly known, which is seldom the case for
astronomical measurements, in particular for complex
astronomical measurements such as caustic masses and velocity dispersions,
whose quoted errors come from a simplified analysis.
Furthermore, no regression method
for a Poisson quantity has been previously published in 
astronomical journals and even
less so for a difference of Poisson deviates.

First of all, because of errors, observed and true values
are not identically equal.
The variables $n200_i$ and $nbkg_i$ represent the true richness 
and the true background galaxy counts in the studied solid angles.
We measured the number of galaxies in both cluster and control field
regions, $obstot_i$ and $obsbkg_i$ respectively, for each of our 53
clusters (i.e. for $i=1,\ldots,53$).
We assume a Poisson likelihood for both and that
all measurements are conditionally independent.
The ratio between the cluster and control field solid angles,
$C_i$, is known exactly. In formulae:
\begin{eqnarray}
obsbkg_i &\sim& \mathcal{P}(nbkg_i) \\
obstot_i &\sim& \mathcal{P}(nbkg_i/C_i+n200_i)
\end{eqnarray}

For each cluster, we have a cluster 
mass measurement and a measurement of the
error associated with this mass, $obslgM200_i$ and $obserrlgM200_i$
respectively.
We assume that the likelihood model is a Gaussian centred on the true 
value of the cluster mass, $lgM200_i$, with a scatter given by the true 
value of the mass error, $\sigma_i$:
\begin{equation}
obslgM200_i \sim \mathcal{N}(lgM200_i,\sigma^2_i)
\label{eqn:eq9} 
\end{equation}

We now need to address the fact that we do not know the true 
value of the mass error and that we only have an estimate of it
i.e. we need to model the relationship
between $\sigma_i$ and $obserrlgM200_i$. 
We use a scaled $\chi^2$ distribution, chosen so that 
$obserrlgM200_i^2$ will be unbiased for $\sigma^2_i$, with the 
(welcome) additional property that positivity is enforced.
\begin{equation}
obserrlgM200_i^2 \sim \sigma^2_i \chi^2_\nu / \nu 
\label{eqn:eqn10}
\end{equation}
Notice that for mathematical reasons we model the relationship between
variances rather than between standard deviations.
The degrees of freedom of the distribution, $\nu$, control
the spread of the distribution, with large $\nu$ meaning that 
quoted errors will be close to true errors.
Our baseline analysis uses $\nu=6$ to quantify 
that we are 95\% confident that quoted errors are correct 
up to a factor of 2 (i.e. that 
$\frac{1}{2}<\frac{obserrlgM200_i}{\sigma_i}<2$, derived via the equivalent
probability statement for $obserrlgM200_i^2$ and $\sigma^2_i$). 
We note that when $\nu=6$, the $\chi^2$ distribution is quite skewed, and
most of the remaining 5\% probability lies below $1/2$.
We anticipate that results are relatively robust to the choice of $\nu$.
The shape of the adopted distribution, a $\chi^2$ distribution, 
is for analogy 
to the case in which the quoted error is derived as a result of 
repeated observations; in such a case, 
standard sampling theory for Gaussian data would have made 
our choice extremely natural.

We now turn to the unobserved quantities in our model. 
for which we will
specify independent prior distributions.
We assume a linear relation between the unobserved mass and $n200$ on the
log scale, with intercept $\alpha+14.5$, slope $\beta$ and
intrinsic scatter $\sigma_{scat}$:
\begin{equation}
lgM200_i \sim
\mathcal{N}(\alpha+14.5+\beta (\log(n200_i)-1.5), \sigma_{scat}^2)
\label{eqn:eqn12}
\end{equation}
Note that $\log(n200)$ is centred at an average value of 1.5 and
$\alpha$ is centred at -14.5, purely for computational advantages in
the MCMC algorithm used to fit the model (it speeds up
convergence, improves chain mixing, etc.).
Please note that the relation is between true values, not
between observed values, which may be biased, 
as we will show in Appendix A for an astronomical sample
affected by Malmquist bias. 

The priors on the slope and the intercept of the regression line in
Equation 9
are taken to be quite flat,
a zero mean Gaussian with very large variance for $\alpha$ and a
Students $t$ distribution with 1 degree of freedom for $\beta$.
The latter choice is made to avoid that properties of galaxy clusters 
depend on astronomers rules to measure angles (from the x or from the y axis).
This agrees with the model choices in
Andreon (2006 and later works) but differs from some previous works
(e.g. Kelly 2007) that instead assume a uniform prior on the
slope $\beta=\tan b$ and, as a consequence, favour some angles 
over others, depending on 
the adopted convention on the way angles are 
measured (i.e. from the x axis counterclockwise as in mathematics, or from the
y axis clockwise as in astronomy). Our $t$ distribution on $\beta$ is 
mathematically equivalent to an uniform prior on the angle $b$.

\begin{eqnarray}
\alpha &\sim& \mathcal{N}(0.0,10^4) \\
\beta &\sim& t_1
\label{eqn:eqn11}
\end{eqnarray}

For the true values of the cluster richness and background,
we have tried not to impose strong a-priori values, only enforcing
positivity.
Both are given independent improper uniform priors.
\begin{eqnarray}
n200_i &\sim& \mathcal{U}(0,\infty) \\
nbkg_i &\sim& \mathcal{U}(0,\infty)
\label{eqn:eqn8}
\end{eqnarray}

Finally we need to specify the prior on the mass error, $\sigma_i$, 
and on the intrinsic scatter of the mass-richness scaling, 
$\sigma_{scat}$. These are positively defined (by definition), 
but otherwise we impose quite weak prior information.
For mathematical reasons, we parameterise these priors on the variance rather
than on the standard deviations as might seem more natural (for astronomers).
An extremely common choice is the Gamma distribution:
\begin{eqnarray}
1/\sigma_i^2 &\sim& \Gamma(\epsilon,\epsilon)\\
1/\sigma_{scat}^2 &\sim& \Gamma(\epsilon,\epsilon)  
\label{eqn:eqn13}
\end{eqnarray}
with $\epsilon$ taken to be a very small number.
The above equations translate almost literally into the 
JAGS code given in Appendix B. The code is only about 15 lines 
long in total, about two orders of magnitude
shorter than any previous implementation of a regression model 
(e.g. Kelly 2007, Andreon 2006), none of which address the noisiness of the
quoted error. 

Our model seems quite complex with a lot of assumptions, 
more than other models
adopted in previous analysis, but actually it makes weaker
assumptions, plainly states what is actually also assumed
by other models (e.g. the conditional independence and
Poisson nature of $obsbkg_i$ and $obstot_i$, the positivity
of the intrinsic scatter, etc.) and
removes approximations adopted in other approaches.
For example, it is common to ignore the uncertainty in the count data and 
to take $n200$ to be the observed $obsn200=obstot - obsbkg/C$.
However, doing so does not respect the fact that $n200$ must be non-negative 
and in the low count regions $obstot - obsbkg/C$ can be
found to be negative (see Appendix B of Andreon et al. 2006).
Instead, we account for the difference and we will show in
the Appendix an example of the danger of ignoring
the difference between $obsn200$ and $n200$. Eq. 5 and 6 also capture
the Poisson nature of galaxy counts that, for small values, is fairly
different from the usual Gaussian approximation widely adopted in
regression models published in astronomical journals.
Furthermore, it is common to ignore the uncertainty in
the mass error. Our model may easily recover this case, by
letting $\nu$ take a large value (formally, to go to infinity). 
Our model replaces this strong assumption with a weaker one, namely that
the quoted squared error is an unbiased measure of the true squared error.
Finally, the remaining ingredients are just uniform (or nearly so)
distributions in the appropriate space.

Essentially, our model assumes that the true richness and true mass are 
linearly
related (with some intrinsic scatter) but rather than having these true 
values we have noisy measurements of both richness and scatter,
with noise amplitude different from point to point.
In the statistics literature, such a model is know as an 
``errors-in-variables regression'' (Dellaportas \& Stephens, 1995).
Our model goes one step beyond 
the works of D'Agostini
(2004), Andreon (2006) and Kelly (2007), which all assume 
errors to be perfectly known (and none of which
deal with Poisson processes as galaxy counts).
These works were, in turn, less approximate approaches
than previous fitting methods used in astronomy to regress two quantities 
(for example, simple least-squares, bivariate correlated error and intrinsic
scatter, etc.).

To summarise, the novelty of the present approach is to treat in a symmetric
way measurements and estimates of errors.
The parameters of primary importance are those of the linear relationship 
between true mass and richness, with associated intrinsic scatter 
$\sigma_{scat}$ being of particular interest. 

\begin{figure}
\psfig{figure=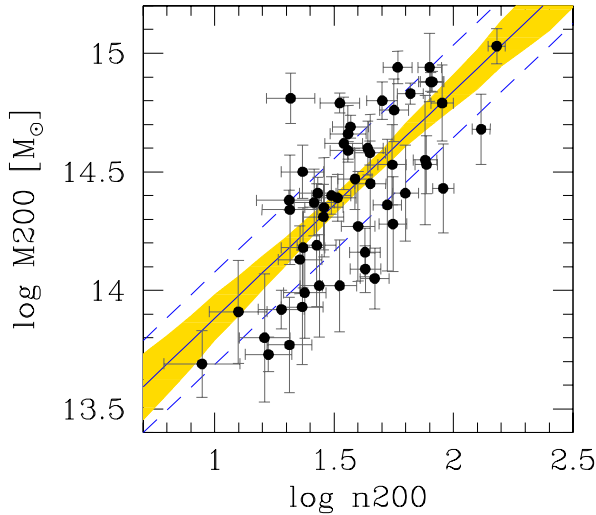,width=8truecm,clip=}
\caption[h]{Richness-mass scaling.
The solid line marks the mean fitted regression line of $lgM200$ on $log(n200)$, while the dashed line
shows this mean plus or minus the intrinsic scatter $\sigma_{scat}$. The shaded region marks the 68\%
highest posterior credible interval for the regression. Error bars on the data points represent observed
errors for both variables. The distances between the data and the regression line is due in part to the
measurement error and in part to the intrinsic scatter.}
\label{fig:fig1}
\end{figure}

\begin{figure*}
\psfig{figure=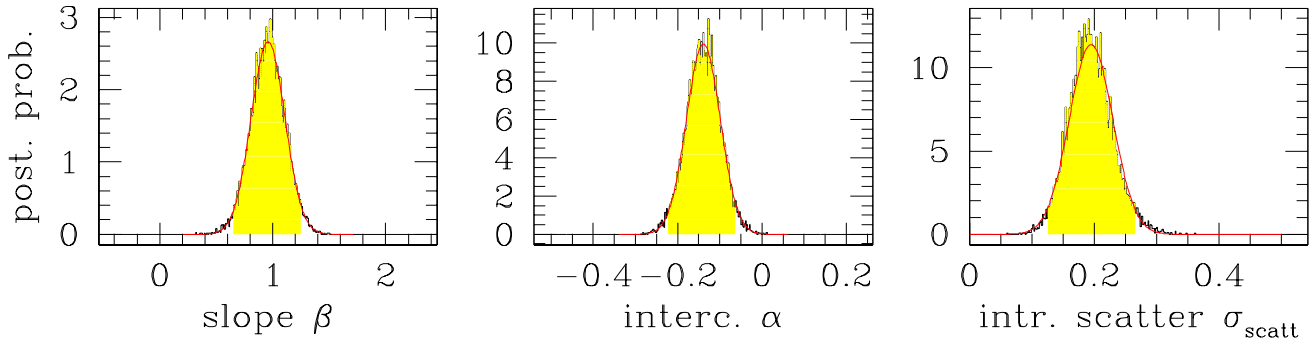,width=18truecm,clip=}
\caption[h]{Posterior probability distribution for the
parameters of the richness-mass scaling.
The black jagged histogram shows the posterior as computed
by MCMC, marginalised over the other parameters. The red curve
is a Gauss
approximation of it.  The shaded (yellow) range shows
the 95\% highest posterior credible interval. 
}
\label{fig:fig2}
\end{figure*}

\subsection{Richness-mass result}

Using the model above, we found, for our sample of 53 clusters:
 
\begin{equation}
lgM200 = (0.96\pm0.15) \ (\log n200 -1.5) +14.36\pm0.04 
\end{equation}
(Unless otherwise stated, results of the statistical computations 
are quoted in the form $x\pm y$ where $x$
is the posterior mean and $y$ is the posterior standard deviation.)

Figure 2 
shows the scaling between richness and mass, observed
data, the mean scaling (solid line) and its 68\% uncertainty (shaded yellow
region) and the mean intrinsic scatter (dashed lines) around
the mean relation.  The $\pm 1$ intrinsic scatter band is 
not expected to contain 68\% of the data points, because of 
the presence of measurement errors. 

Figure 3 
shows the posterior marginals for
the key parameters; for the intercept, slope and intrinsic
scatter $\sigma_{scat}$. These marginals
are reasonably well approximated by Gaussians.
The intrinsic mass scatter at a given richness, 
$\sigma_{scat}=\sigma_{lgM200|\log n200}$, is small, $0.19\pm0.03$.
The small scatter and its small uncertainty is promising from the point
of view of
using $n200$ for cosmological aims, for example to estimate
the mass distribution, given the $obsn200$ distribution.

The slope between richness and mass 
is very near to one (within one third of the estimated standard deviation), i.e.
clusters which have twice as many galaxies are twice as massive.

\begin{figure}
\psfig{figure=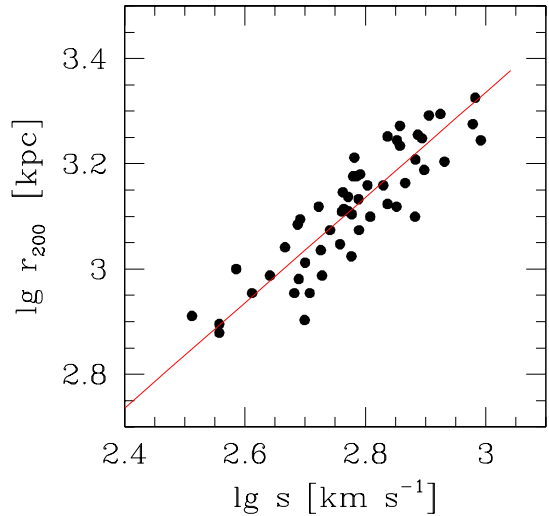,width=8truecm,clip=}
\caption[h]{$r_{200}-s$ (velocity dispersion) scaling.
The line marks the expected scaling, $r_{200} \propto s$. The
good agreement between the trend identified by the data and the
expected scaling implies that there is no velocity dispersion (mass) dependent
systematic bias on the adopted $r_{200}$.}
\end{figure}

\subsection{Checks}

Firstly our results are robust to the choice
of $\nu$ (we tested $\nu=6$ vs $\nu=3,30,300$ or $\nu=3000$). 

Second, the determination of the slope of the richness scaling requires setting
two (astronomical) parameters, a radius within which galaxies should be counted
and a limiting (reference) magnitude.
To investigate the dependence of the richness-mass slope on which 
limiting magnitude is adopted, we recompute $n200$
using two different limiting magnitudes, one and two mag deeper than
our reference mag, $M_V=-20$ mag. 
The resulting slopes of the mass-richness scaling are $0.98\pm0.15$
and $0.95\pm0.16$, both very close to the original slope 
derived using the reference mag ($0.96\pm0.15$).
The intrinsic scatter changes insignificantly, by 0.01 dex, with
the limiting magnitude. 

We now check whether the scaling of richness found with mass 
may be biased (tilted) by having 
hypothetically taken a systematically incorrect $r_{200}$ (for example, 
too small an $r_{200}$ at large masses, or too big a one at small masses).
Figure 4 plots $r_{200}$ as a function of
cluster velocity dispersion. 
The superimposed straight line comes from assuming that 
$r_{200}$ is the virial radius (i.e. $M_{200}=M_{virial}$), 
$r_{200} \propto s$ (e.g. eq. 1
in Andreon et al. 2005, eq. 1 of Carlberg et al. 1997, eq. 3.1 in 
Muzzin et al.
2007) rather than as a fit to these points.
As the points are scattered roughly around the slope of 
the expected relation, we reject the possibility that the slope
between richness and mass (or velocity dispersion) is biased because of
a bad choice of the reference radius in which
galaxies are counted (one that does not correctly scale with mass).

In summary, $n200$ tightly correlates with 
mass, with 0.19 dex intrinsic scatter. The 
slope is fairly robust to the choice of the reference magnitude,
the uncertainty of error terms ($\nu$) and the a-priori range of 
mass errors. Furthermore, it is unbiased with respect to a (hypothetical) 
bad choice of the reference radius.

\begin{figure}
\psfig{figure=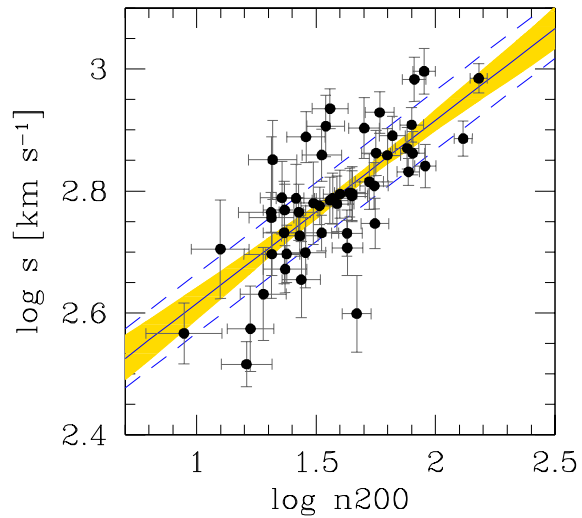,width=8truecm,clip=}
\caption[h]{Velocity dispersion -- richness scaling.
Symbols are as in Fig. 2.
}
\end{figure}

\subsection{Richness-velocity dispersion scaling and results}

Velocity dispersions, $s$, are observationally more expensive
than $n200$ but less expensive
than caustic masses. They are also good tracers of
the cluster mass (e.g. Biviano et al. 2006; Mandelbaum \& Seljak
2007; Evrard et al. 2008). Since at high redshifts 
caustic masses are observationally prohibitive to calculate,
from the perspective of testing the evolution
of the richness scaling
it is useful to
calibrate the scaling between richness and velocity dispersion.

The  statistical model employed is very similar to that described for
the richness-mass scaling, essentially we
only need to read ``velocity dispersion" where mass was
written. Because velocity dispersion errors are easier to measure
than mass errors, we adopt $\nu=50$, i.e.
we are 68\% confident that quoted errors are correct up to a
factor 1.1 (i.e. within 10\%).

Because of the different measurement units, the intercept $\alpha$
is now centred at $2.8$ (for computational purposes in 
JAGS).  

For our sample, we found 
\begin{equation}
\log s = (0.30\pm0.04) \ (\log n200 -1.5) +2.77\pm0.01 
\end{equation}

Figure 5 shows the fitted scaling between richness and velocity dispersion, the 
observed data, the posterior mean scaling (solid line) and its uncertainty
(shaded yellow region) and the mean intrinsic scatter (dashed lines).

Similarly to the richness-mass scaling, the
intercept, slope and intrinsic scatter
have posterior marginals which are close to Gaussian.
The intrinsic  velocity dispersion scatter at a given richness, 
$\sigma_{scatt}=\sigma_{\log s |\log n200}$, is small, $0.05\pm0.01$.

As in the case of the richness-mass scaling, these results
are robust to the choice of $\nu$, for $\nu \ga 10$.

The fitted slope of the richness -- velocity dispersion scaling is one third
of the slope of the richness -- mass scaling, as it should be, given that
velocity dispersion scales with mass with power 0.33 (e.g. Evrard et al. 
2008).

\begin{figure}
\psfig{figure=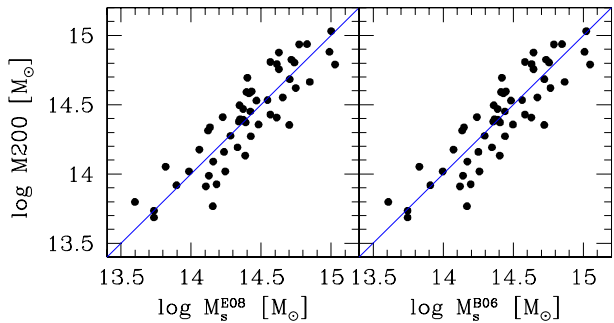,width=8.5truecm,clip=}
\caption[h]{Caustic masses (ordinate) vs masses derived from 
the cluster velocity
dispersion using relations calibrated with numerical simulations
(left panel: Evrard et al. 2008, right panel: Biviano et al. 2006).
The solid, slanted, line mark the equality and it is not a fit
to the data. 
}
\end{figure}

\subsection{Caustic mass systematic errors}

In the previous sections we have not accounted for possible systematic 
error
in the caustic mass, except indirectly in a couple of locations:
a) in Figure 4, when comparing
the cluster velocity dispersion with $r_{200}$: if a systematic
error
on M200 were present, then the data would not scatter around
the expected relation; b) in section 5.4, where we found
that the slope of richness-velocity dispersion is one third of
the slope of the richness-mass, as expected for a mass that
scales with the cube of velocity dispersion.

In order to further investigate the lack of gross systematic errors on
caustic masses we plot in Figure 6 caustic masses 
against two masses, derived from velocity
dispersion using relations calibrated with numerical simulations
(left panel: Evrard et al. 2008, right panel: Biviano et al. 2006). 
The solid line is the one-to-one relation
rather than a fit to the points. If caustic masses were
systematically larger or smaller than masses derived
from velocity dispersion, then these points might well be
systematically above or below the solid line. If instead
caustic masses were too big at high masses and too small at
low masses, or vice versa, 
points should have a different (tilted) slope from the plotted
line. Figure 6 clearly shows that neither of the two cases occurs.
A 30 \% offset error or a 30 \% tilt would be obvious to the eye.
A second obvious conclusions coming from this figure is
that the two panels are virtually indistinguishable. 
This is because the two calibrations of the velocity 
dispersion-mass relation, although independent, are 
actually almost identical. 

To summarise, this section shows the lack of
an obvious gross systematic error in caustic masses. 
``Statistical" errors on caustic mass and noisiness
of errors are built-in in our model.

\section{Richness as mass proxy}

The richness-mass scaling derived in previous sections
needs a known $r_{200}$,
the radius within which galaxies have to be counted.
If we want to use $n200$ as a mass proxy, $r_{200}$ should
be instead considered as unknown.  Lopes et al. (2009)
disagree with this reasoning because in their
work they measured the performance of
mass proxies assuming $r_{200}$ (or $r_{500}$) known, when
instead it is unknown for clusters with unknown masses.
We now measure the performances of a richness estimate
that does not require the knowledge of $r_{200}$, counting 
galaxies within some reference radius, $\widehat{r_{200}}$,
that can be measured from imaging data\footnote{The hat above symbols
is introduced to distinguish these values, derived from
eq. 18, from values used in previous sections which were taken from
CIRS.}. Since
there are a number of ways $\widehat{r_{200}}$ may be
estimated, we consider some of them.

In principle, we may be interested in:

a) $\sigma_{scat}$, i.e. the intrinsic scatter in mass at a given richness. This may be
of interest to those who want to known which part of the
observed scatter is intrinsic. 

b) the uncertainty of the mass estimated
from the cluster richness. This is,
for example, the case when one has one or few clusters with a measurement of
richness and we would like to know their estimated mass.
With real data, cluster richness is known with a finite precision
which induces a minimal floor in the performances
of richness as mass proxy.

To this end, we first need to find a way to estimate $\widehat{r_{200}}$
from galaxy counts, because clusters for which we want an estimate of
mass will not have a known $r_{200}$. Then we will
calibrate the measured $\widehat{n_{200}}$
($n_{200}$ values within $\widehat{r_{200}}$) with mass and
estimate the uncertainty of the predicted
$lgM200$ for a cluster sample, the latter using Eq. 4.
Recall that the
performance of richness as a mass predictor
accounts for
all terms entering into the modelling of proxy and mass, 
which include the uncertainty of the proxy value and 
the uncertainty on the parameters
describing the regression between mass and mass proxy (slope, intercept, 
intrinsic scatter and their covariance).

As in some literature approaches, we use the same sample 
both to establish the scaling between regressed
quantities and to measure the proxy performance. 
However, these literature
approaches compute the proxy performances 
from a single regression (usually
named the best fit, i.e. for a single value of $\theta$), ignoring that 
other fits are similarly acceptable and that the best fit itself  
is uncertain (i.e. ignoring
uncertainties on slope, intercept and intrinsic scatter). 
When the best fit scaling is defined as the
one minimising the scatter (and this is not our case), 
the measured scatter
underestimates the true scatter, by definition.
Our approach supersedes these previous approaches,
allowing for errors other approaches neglect and also
including their covariance.

\begin{figure}
\psfig{figure=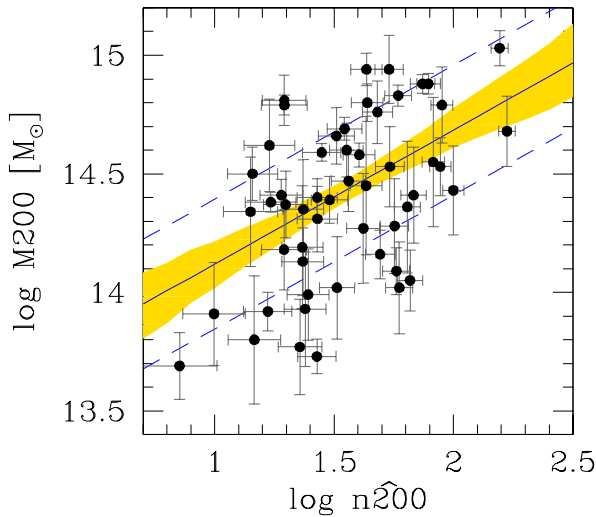,width=8truecm,clip=}
\caption[h]{Richness--mass scaling 
for a richness measured within $\widehat{r_{200}}$, an $r_{200}$ radius
estimated from optical data. Symbols are as in Fig 2.
}
\end{figure}

\subsection{Reference case}

Since we do not known a priori which approach is the optimal
way to estimate $r_{200}$ from imaging data alone, in this
section we consider a reference case and in the following section we
make a number of tests to see how robust are our conclusions to
the assumptions made in the reference case.

We simply compute the number of cluster galaxies (i.e. $obstot-obsbkg/C$)
within a radius of 1.43 Mpc, $obsn(r<1.43)$ and then we estimate 
$\widehat{r_{200}}$ as 

\begin{equation}
\widehat{\log r_{200}} = 0.6 \ (\log obsn(r<1.43)-1.5)
\end{equation}

The slope, $0.6$ and the radius, $1.43$ Mpc are taken for 
consistency with Koester et al. (2007). The intercept
is chosen to reproduce the trend between known $obsn(r<1.43)$ 
and $r_{200}$ radii. Therefore, our $\widehat{r_{200}}$ has no 
bias (or at most a negligible one) with respect to
$r_{200}$ by construction. Instead,
the normalisation (intercept) adopted in Koester et al. (2007) has
been later discovered (Becker et al. 2007; Johnston et al. 2007) 
to give radii too large by a factor of two. 

Having adopted the radius above, we need to count the galaxies
within this radius and recompute the
solid angle ratio.
Asymptotically, $\widehat{n200}$ is given by
$\widehat{obstot}-obsbkg/\widehat{C}$, but our analysis does not 
assume that this asymptotic behaviour holds within our finite
sample\footnote{Background counts do not need to be
recomputed, which is why there is no hat on $obsbkg$.}.

Using our fitting model, we found 

\begin{equation}
lgM200 = (0.57\pm0.15) \ (\log \widehat{n200} -1.5) +14.40\pm0.05 
\end{equation}
The data and fit are depicted in Figure 7.
The major difference with respect to our fit performed
on measurements using knowledge of $r_{200}$ (i.e. sec 6.2) is
the shallower slope, which is 1.9 combined $\sigma$ shallower
than it was. 
Intercept, slope and intrinsic scatter have posteriors close to Gaussian. 
The intrinsic scatter
is small, it has mean $0.27\pm0.03$ dex. 
With respect
to the case where $r_{200}$ is known, the intrinsic scatter
is larger ($0.27\pm0.03$ vs $0.19\pm0.03$), as
expected because we are not using 
our knowledge about $r_{200}$. We emphasise
that this is the uncertainty on the mass inferred from $\widehat{n200}$
if we were able to measure the latter quantity with very large precision,
being $0.27$ dex the part of the mass scatter not associated to 
measurement errors.
Since $\widehat{n200}$ is not better known than allowed
by the observed data, the mass error inferred
from a (noisy) estimate of the cluster richness is larger and is given by
the average uncertainty of predicted $lgM200$, which is
found to be $0.29\pm0.01$ dex. Therefore, we can predict the mass 
of a cluster within $0.29$ dex by measuring
its richness. 
Since the uncertainty on the predicted mass is only
slightly larger than the intrinsic scatter, the uncertainty
on the mass-richness scaling (regression) and proxy uncertainty
only account for
small amounts of the variability. Therefore the performance of richness as mass
proxy is dominated by the mass scatter at a given richness.

In comparison to caustic cluster masses, which have, on average, 
a $0.14$ dex error, masses estimated from $\widehat{n200}$ have
twice worse accuracy ($0.29$ vs $0.14$ dex).
Although $\widehat{n200}$ is noisier mass proxy than caustic
masses, the former requires far less expensive observations 
than the latter and as a consequence 
is available for almost a two hundred times larger sample. 
The last number is computed
as the ratio of the number of clusters with available richness
from SDSS (e.g. maxBCG clusters, about
13000, Koester et al. 2007) 
with those with caustic masses
in the same sky region (74, listed in Rines \& Diaferio 2006).

\subsection{Other paths to $\widehat{n200}$}

\begin{figure*}
\psfig{figure= 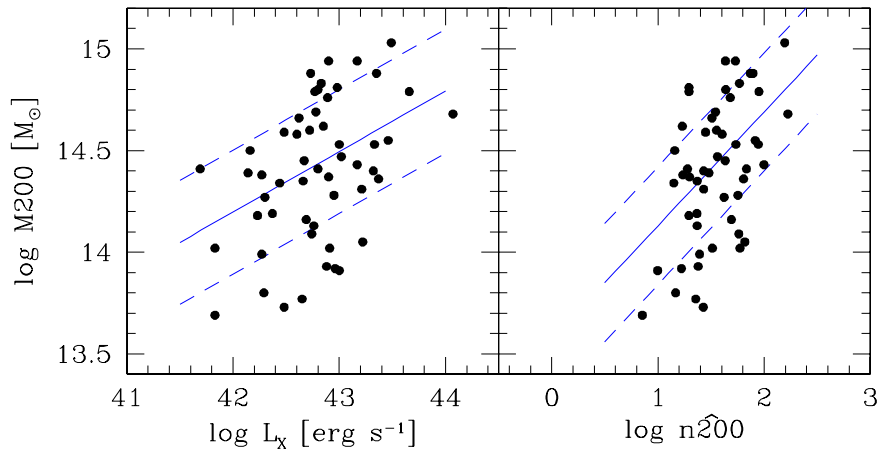,width=12truecm,clip=}
\caption[h]{Comparison of the performances, as mass predictor, of
X-ray luminosity and richness. The solid line mark the mean model,
the dashed lines delimit the mean model plus and minus the average
uncertainty of predicted masses. Equal ranges (3.5 dex) are adopted
for richness and X-ray luminosity.  See section 4 for a discussion of 
the slopes of prediction lines.
}
\end{figure*}

To check the resulting robustness of our results to the few parameters
involved in the computation, we make some tests.
First, we take as reference radius a value near
to the average of our $r_{200}$, 1.25 Mpc and use $obsn(<1.25)$ as pivot
value for estimating $\widehat{r_{200}}$. Second, we changed the
slope to $0.55$, because some maxBCG papers (Koester et al. 2007; 
Hansen et al. 2005; Becker et al. 2007) disagree on the 
slope value ($0.55$ or $0.6$) adopted in Koester et al. (2007).
Third, we decide to count galaxies in a radius twice larger than
$\widehat{r_{200}}$, to check the sensitivity to the adopted
reference radius. The factor two is adopted to follow the maxBCG papers,
which adopted an $r_{200}$ radius later discovered to be too large by a
factor of two.
Fourth, we consider the simplest case, we adopt a fixed
aperture, 1.43 Mpc, for all clusters, irrespective of their
size or mass. In all these cases we found similar
slopes, intrinsic mass scatter and average uncertainty of predicted $lgM200$ 
as in our reference case. This is expected,
given that the intrinsic scatter alone accounts for most of
the uncertainty of predicted masses.

In summary, if $r_{200}$ has to be estimated from a scaling relation based
on counting red galaxies within in aperture in imaging
data, it seems that we have reached a floor on the quality of
mass determination, $0.27$ dex of intrinsic scatter,
and $0.29$ dex of average  
uncertainty of predicted $lgM200$,
no matter how precisely $\widehat{r_{200}}$ is defined.

\subsection{Comparison with other mass proxies}

In this section we want to compare the performances of the X-ray
luminosity and richness as mass proxies. Among all possible
proxies, we choose X-ray luminosity because it is measurable from survey
data, as is richness. Other mass proxies, such as $Y_X$, do require
follow-up observations and it would be unwise to compare them to
(optical) mass estimates derived from survey data.
Of course, in this comparison, both mass proxies
are measured without using knowledge of mass or
linked quantities, such as $r_{200}$, because they are
unknown for clusters with unknown masses.  Lopes et al. (2009)
disagree with this reasoning 
because they measured and compared the performance of
mass proxies assuming knowledge of $r_{200}$ (or $r_{500}$).

Richness and its performance
as a mass predictor have been measured by us in the previous section.
In short, richness offers a mass with a $0.29$ dex uncertainty.
X-ray luminosities are collected by Rines \& Diaferio (2006) and 
come, in order of preference, from the ROSAT-ESO Flux-Limited X-Ray
(REFLEX), the Northern ROSAT All-Sky galaxy cluster
survey (NORAS), the Bright
Cluster Survey (BCS) and its extension (eBCS) and, finally, from the 
X-Ray Brightest Abell Cluster Survey (XBACS).
Rines \& Diaferio (2006) do not list errors for
X-ray luminosities,
therefore we repeat our analysis with 
a 5\% and a 30\% error and we found that
results are robust to the adopted error.
The model
for the logarithm of the X-ray flux is assumed to be Gaussian and
the following equations

\begin{eqnarray}
obslgLx_i & \sim & \mathcal{N}(lgLx_i, err^2) \\
lgLx_i & \sim & \mathcal{U}(0,\infty) \\
lgM200_i & \sim & \mathcal{N}(\alpha+14.5+\beta (lgLx_i-42.5), \sigma_{scat}^2) 
\end{eqnarray}
replace eq. 5, 6 and 9.  Before proceeding further, we emphasise
that our analysis involving $L_X$ ignores the Malmquist bias due
to the X-ray selection of the cluster sample (e.g. Stanek et al. 2006),
i.e. clusters brighter than average for their mass
are over-represented (easier
to detect and thus more likely to be in the sample).

Figure 8 shows richness vs mass and X-ray luminosity vs mass, 
the fitted scaling (posterior mean, solid line) and the mean
model plus and minus the uncertainty of predicted masses (dashed lines).

By eye, our fit seems shallower than the data suggest.
Our derived slopes match those derived by other fitting algorithms, 
for example, the $L_X$ vs mass regression has 
a slope of $0.30\pm0.10$ using our fitting
algorithm, a slope of $0.29\pm0.10$ neglecting the uncertainty
on the error (i.e. following Andreon 2006 and
Kelly 2007) and a BCES$(Y|X)$ (Akritas \& Bershady 1996)
slope of $0.31\pm0.07$. This slope is
theoretically different from the slope of the underlying
relation between these quantities 
because we are interested here in something
different, namely prediction as explained in Sec 4. 
As a further cautionary check,
we verified that the uncertainty of predicted masses, the
quantity of interest here, is robust, in particular
we forced a steeper slope (e.g. we keep the $L_X$-mass slope to $0.5$),
getting an identical value for the uncertainty.

For the richness, we found (sec 7.1)
a mass uncertainty of $0.29\pm0.01$ dex.
For the $L_X$ proxy, we found an identical value for the
mass uncertainty,  
$0.30\pm0.01$ dex. 
Therefore,
masses predicted by $L_X$ or richness are  comparably precise, to 
about $0.30$ dex.  Qualitatively, one may reach the same
conclusion by inspecting Figure 8 and performing an approximate
analysis requiring a number of assumptions
that are unnecessary in our statistical analysis; the precision
of a mass proxy is, in our case, dominated by the intrinsic scatter
in mass at a given proxy value, which in turn is not
too dissimilar from the vertical scatter in Figure 8 because observational
mass errors are not large. The two data point clouds display similar widths 
at a given value of the proxy (see Figure 8)
and therefore the two proxies display similar
performances as mass predictors. Our statistical analysis removes
approximations and holds when the qualitative
analysis does not, for example if the regression is poorly determined,
or the mass errors are large, or the richness is poorly determined,
or in the presence of a mismatch in proxy value between clusters with
known and to-be-estimated mass.

A 
plot similar to our Figure 8 by Borgani \& Guzzo (2001) 
seems to show a better $L_X$ performance,
but only when compared to an optical richness estimated by eye (Abell 1958; 
Abell et al. 1989).

\begin{figure*}
\psfig{figure= 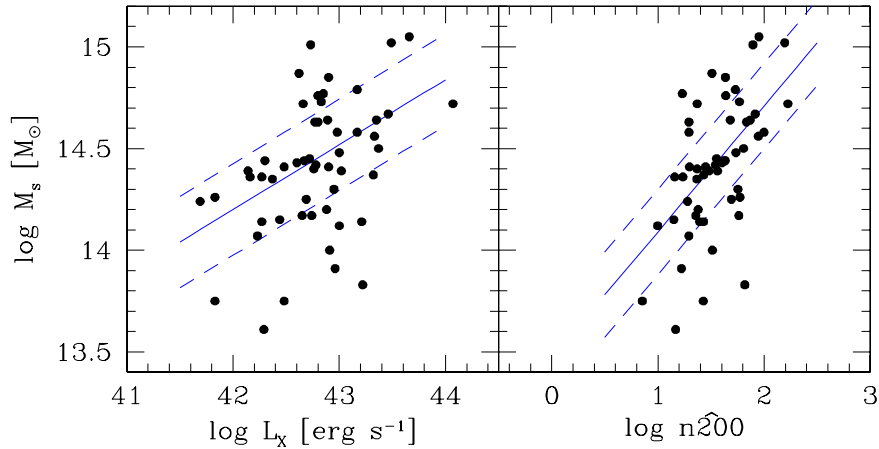,width=12truecm,clip=}
\caption[h]{Comparison of the performances, as mass predictors, of
X-ray luminosity and richness. In this figure we use masses inferred
from cluster velocity dispersion, but the basic result does not change,
X-ray luminosity and richness score similarly as mass proxies. 
Lines and symbols as in previous
figure.
}
\end{figure*}

The fact that the studied sample is mainly (but not exactly), 
an X-ray flux-limited sample gives an advantage
to $L_X$ as a mass proxy; had we taken a volume complete mass-limited
sample of the same cardinality in the same Universe volume 
(i.e. unbiased with mass) instead of the
adopted (almost) flux-limited sample, some clusters would not be X-ray
detected and thus would have a very loose mass constraint, lacking an
$L_X$ detection. A richness-selected sample formed by
all clusters with $n200$ above a threshold would also have favoured
the $n200$ mass proxy, because the scatter between
$n200$ and $L_X$ would have included in the sample
clusters undetected in X-ray. Therefore, 
in spite of the selection favouring the X-ray proxy, richness  
performs as $L_X$ in predicting cluster masses.

Richness has a further advantage, it is available for a larger
number of clusters per unit Universe volume.  Let us consider for example
$z<0.3$, the number of clusters with optical estimates of mass (i.e.
with $\widehat{n200}$) outnumber the one with an X-ray based proxy
(i.e. with $L_X$) by a factor 58; there are 5800 ster$^{-1}$ optically
detected clusters (maxBCG clusters, Koester et al. 2007) and only 100
ster$^{-1}$ X-ray detected (REFLEX clusters, Bohringer et al. 2001).
The optically-selected cluster sample is a quasi volume limited sample.
If,  mimicking what has been done 
for X-ray measurements,
a cut on $\widehat{n200}$ signal to noise is used, instead of adopting
a quasi volume limited sample,
the number of optically detected clusters grows significantly.
Similarly, in about four degrees squares, there are about 
106 clusters with $obsn200>6$ (Andreon et
al., in preparation) and $0.32<z<0.8$. In the very same
area there are 9 C1 clusters (Pacaud et al. 2007), i.e.
ten times fewer. If, as for X-ray data, a
cut on $obsn200$ signal to noise is adopted, the number
of optically detected clusters would be larger. Schuecker, Bohringer and
Voges (2004) claim SDSS being deeper than the Rosat All Sky Survey
(RASS) ``can thus be used to guide a cluster detection in RASS
down to lower X-ray flux limits". Unsurprisingly, the fact that 
COSMOS (Scoville et al. 2007) and many X-ray cluster surveys keep
X-ray detections that match with optical clusters (e.g. 
Finoguenov et al. 2007, for COSMOS) {\it assumes} that in current data
sets X-ray clusters are a sub-set of optically selected clusters, i.e.
a smaller sample.  Finally, while X-ray
selected clusters are almost always optically detected, the reverse
has proved much more difficult, which clarifies that optical cluster detection 
is observationally 
cheaper. Therefore, studies that require large samples of
clusters or a denser sampling of the universe volume may adopt 
optically-selected cluster samples because they offer a 
mass estimates of comparable quality for larger samples.

Some words of caution are in order. The good performance of richness as
a mass estimator holds for our sample and should be confirmed on a sample of
clusters optically selected. Of particular relevance is
the frequency of catalogued optically selected clusters
being instead line of sight superpositions of smaller systems.
Such points will be addressed by our X-ray follow-up of
all (53) clusters optically selected
with $59<n_{200}<70$ and $0.1<z<0.3$ in the maxBCG catalogue
(Koester et al. 2007a).

Similarly, the same caution is in order for other 
mass estimators. For example $L_X$ has
been proposed as a mass estimator by Maughan (2007). 
Its performances as a mass predictor, however, has been measured on
data having $L_X$
based on hundreds or thousands of photons and therefore the
noisiness of $L_X$  itself in establishing his
performances as a mass estimator has been largely underestimated.
Furthermore, point sources are identified and
removed through (high-resolution) Chandra observations, 
making the identification and 
flagging of point sources easy and studied
clusters have preferentially large count rates and are
little affected by residual, unrecognised as such, 
point sources. To summarise, the good
performances of $L_X$ cannot be immediately extrapolated to common
cluster samples, dominated by objects with noisy $L_X$/count rate 
(because of the steep cluster number counts) and 
for which residual point source contamination is
more important and which are perhaps observed by survey
instruments as XMM, having a lower resolutions and
therefore a more difficult identification and of
contaminating point sources.

\section{A third mass calibration}

Our approach can be used to calibrate richness
against mass, no matter which mass we are talking about
(e.g. lensing, caustic, Jean, etc). 
In section 6.4 we use velocity
dispersion (uncorrected with any numerical simulation) to
calibrate the richness scaling, recycling the 
same model already used for caustic masses. 
As a further example,
we recycle our model to calibrate richness against $M_s$, the
mass derived
from velocity dispersion, $s$, fixed with a mass-$s$ 
relation derived by numerical simulations. We
adopt the mass-$s$ relation in Biviano et al. (2006). 
As shown in Figure 6, had we used the mass-$s$ relation 
in Evrard et al. (2007) we would have found near indistinguishable 
results.
To use the masses $M_s$ in place of the caustic ones, 
we need only write their values (and their errors) in the data file and 
run our same model. Mass errors are derived
by combining in quadrature velocity dispersion errors (converted
in mass) and 
the intrinsic noisiness of $M_s$ (12 \%, from Biviano et al. 2006).
We adopt, as for
velocity dispersions, $\nu=50$ (but results do not
depend on $\nu$, if $\nu\ga 30$). 
We found, for our sample of 53 clusters:

$ lgM_s = (0.92\pm0.11) \ (\log n200 -1.5) +14.35\pm0.03 $

with an intrinsic scatter of $0.12\pm0.04$ dex.
The intercept, slope and intrinsic scatter
have posterior marginals which are close to Gaussian,
as for the scaling with caustic masses.

Unsurprisingly, we found a near identical slope and intercept
to those using caustic masses; to find different values we would
 need
that caustic masses be tilted or offset from 
velocity-dispersion derived masses, whereas Figure 6 shows the
lack of a gross tilt or offset.

We also found compatible values for intrinsic scatter 
($0.12\pm0.04$ vs $0.19\pm0.03$). The similarity of the
two intrinsic scatters testifies that
errors on caustic masses and on $M_s$ are on a consistent
scale, i.e. similarly correct (or incorrect); for example, if
the caustic mass error is overestimated, the intrinsic
scatter of the caustic mass-richness scaling would be lower
than the one using $M_s$, because the intrinsic scatter
is the part of the scatter not accounted for the measurement
errors.

We now move on to consider $\widehat{n200}$, the 
cluster richness estimated without knowledge of the cluster mass
and linked quantities, as $r_{200}$. How does it perform
as a mass proxy, when $lgM_s$ is used as mass? At the minimal 
effort
of listing the data in the data file, we find:

$lgM_s = (0.62\pm0.12) \ (\log \widehat{n200} -1.5) +14.40\pm0.04 $

\noindent
ie, indistinguishable from the scaling with caustic masses,

$lgM200 = (0.57\pm0.15) \ (\log \widehat{n200} -1.5) +14.40\pm0.05 $
The intercept, slope and intrinsic scatter have posteriors close to Gaussian,
as with the scaling with caustic masses. The intrinsic scatter
is small, it has mean $0.21\pm0.03$ dex, very similar to the one
obtained using caustic masses, $0.27\pm0.03$ dex.  
The average uncertainty of predicted $lgM_s$, i.e. the quality of
richness as mass proxy, is
found to be $0.23\pm0.01$, similar to the one obtained using
caustic masses, $0.29\pm0.01$ dex. 

To explore the quality of $L_X$ as a mass proxy when mass is measured
by $lgM_s$, we only need to
list the data and run our model, with no change. We found
that mass, predicted from $L_X$ has an average uncertainty of
$0.22\pm0.01$. When caustic masses are used,
the average uncertainty of predicted masses was $0.30\pm0.01$. 

Figure 9 shows the richness vs mass
and X-ray luminosity vs mass, for velocity-dispersion derived masses.
It is the equivalent of Figure 8.
The bottom line is hardly different
from that derived in sec 7.3;
richness and X-ray luminosity show comparable performances in
predicting cluster mass (either caustic or derived from cluster
velocity dispersion). If anything, there is some tentative evidence
that both X-ray luminosity and richness better predict
masses derived from velocity dispersions 
than caustic masses ($\sim0.21$ vs $\sim0.29$). We will defer to a 
future paper an in-dept examination of the significance of this
possible effect.

\section{Discussion and Conclusions}

In order to exploit clusters as cosmological probes, it is important
to know the mass-proxy scaling. Although self-solving for
the scaling itself is feasible, an independent
calibration of the scaling
is a safety check and allows us to improve cosmological
constraints.

In this paper we computed the richness (number of red galaxies
brighter than $M_V=-20$ mag) of 53 clusters with
available caustic masses, the latter
having the advantage that, unlike other masses, 
they do not require the cluster to be in 
dynamical or hydrostatic equilibrium.  We investigated the
possibility of systematic biases by comparing caustic masses 
to masses derived from velocity dispersions and found
no gross offsets or tilt.
Richness is computed from SDSS imaging data both with and without 
knowledge of the reference radius $r_{200}$ from SDSS imaging
data. We then measure the scaling between richness and caustic mass.
Our richness-mass calibration is solid, both from an astrophysical
perspective, because we the adopted masses are amongst
the most hypothesis-parsimonious estimates of cluster
mass and statistically, because we account 
for terms usually neglected,
such as the Poisson nature of galaxy counts, the 
intrinsic scatter and uncertain
errors. Our cluster sample is
larger, by a factor of a few, than previous samples used
in comparable works. The data and code used for the stochastic computation
are distributed with this 
paper.  This code is quite general, we used it to derive two
alternative richness-mass calibrations, using as a mass proxy 
the cluster velocity dispersion or a mass calibrated from the
velocity dispersion via numerical simulations. 

We found a slope between richness and (caustic) mass of $0.96\pm0.15$ with
knowledge of $r_{200}$, i.e. clusters which have twice the number of galaxies are twice as massive.
The intrinsic scatter is small, $0.19$ dex. 
 An identical result is found using masses calibrated from the
velocity dispersion via numerical simulations.  
When the reference radius in which galaxies should be
counted has to be estimated from optical data,  the slope
decreases to $0.57\pm0.15$ and masses inferred
by the cluster richness are good to within $0.29\pm0.02$ dex, 
largely independently of the way the radius itself is 
estimated.
The uncertainty of predicted masses 
is twice the average uncertainty of caustic masses (0.14 dex), but
observationally less expensive to obtain and for this reason
available for a two hundred
times larger sample. Richness is a  
mass proxy of quality comparable to X-ray luminosity, both
showing a $0.29$ dex mass uncertainty, 
but is less observationally expensive
than the latter, as testified by the larger number
density of optically-detected clusters with respect to X-ray
detected clusters in current catalogues. 
This has important applications in the
estimation of cosmological parameters from optical cluster surveys, because in
current surveys clusters detected in the optical range outnumber, by at least one
order of magnitude, those detected in X-ray.
In particular, we note
that our richness is computed 
from the shallowest data ever used
by us, 54 s exposures at a 2m telescope, taken
under mediocre seeing conditions (1.5 arcsec FWHM), i.e. SDSS
imaging data.  Similar or better data should be available for
every cluster; we are unaware of a cluster of galaxy claimed to be
so without some optical imaging of it.

 The similar performances of X-ray luminosity and richness
in predicting cluster masses has been confirmed using cluster
masses derived from velocity dispersion fixed by numerical 
simulations.

People wanting to estimate the mass of one or two clusters
have to measure galaxy counts brighter than $M_V=-20$ mag
within the $r_{200}$ radius estimated from eq. 18 plus a similar measurement
in an area devoid of cluster galaxies to account for background galaxies, 
list these values with our measurements
and run the JAGS code listed in the appendix. Those in a hurry
and accepting a reduced quality of the mass estimation and of its uncertainty
may simply insert the measured $n200$ in eq. 19 and take a 
$\pm 0.29$ dex mass error.

 In the appendix we present an individual comparison with
the literature addressing the richness-mass scaling.
Here we emphasise that our measurement
of the performances of mass proxies conceptually differs from
some other published works; a) we quote the posterior predictive uncertainty
and not the scatter. The former accounts for the uncertainty
in the richness-mass scaling, while the latter does not. Since the
scaling between mass and richness is not known perfectly, we
prefer posterior predictive uncertainty to the scatter. b) Our
own measurement of the scatter is not biased low, whereas literature
values are sometimes biased low as a result of the way the best fit
model is found, minimising the scatter. The best fit relation
is preferred (by other authors) to the true relation if this leads 
to a lower scatter.
The effect is intuitively obvious (and quantitatively important) 
for small samples. We prefer, instead, not to be optimistic. 
c) Some works (e.g. Lopes et al. 2009) evaluate the performances
of a mass proxy assuming that mass-linked quantities, such as $r_{200}$
are known, while they are unknown for clusters with unknown
masses. This logical inconsistency has an important impact on
the final result. Had we followed Lopes et al. (2009), we should have
concluded that richness returns masses with a 0.19 dex precision, instead
of with a 0.27 dex precision, almost a 50 \% underestimate. d) Some works, instead,
forget important items, such as Malmquist bias.
As detailed in the Appendix where individual works are considered, generally
speaking, authors tend to be more optimistic about 
the quality of the richness-mass calibration and
the proxy performances than their data allow.

As mentioned, in order to use richness for cosmological
studies, we need to check that our results hold for
an optically selected cluster and, if a large redshift
range is considered, we need to measure the evolution of
the scaling, similarly to what is necessary for calibrating 
every other mass proxy. The first issue will be 
attacked by our (running) X-ray observations of an optically
selected cluster sample, the second one by a lensing analysis
of an intermediate redshift ($0.3<z<0.8$) cluster sample.
From this perspective, in this paper we also calibrated richness against 
cluster velocity dispersion, which are easier to measure than caustic
masses.
Evolution of red galaxies is now well
understood (Stanford et al. 1998; Kodama et
al. 1998; De Propris et al. 1999; Andreon 2006; Andreon et
al. 2008b), quite differently 
from another
widely used mass tracer, the X-ray emission from the intracluster
medium. For the latter 
one is forced to assume self-similar evolution for lack of
better knowledge (e.g. Vikhlinin et al. 2009) even if available X-ray 
observations argue against 
this scenario (e.g. Markevitch 1998, Pratt et al. 2008).

\section*{Acknowledgements}

We thank Vincent Eke and the anonymous referee 
for their useful comments
which prompted us to insert Sec 3, 6.5 and 8. We also thank
D. Johnston, A. Moretti, R. Trotta, L. Aguillar and E. Meyer 
for useful suggestions.
For the standard SDSS acknowledgement see, 
http://www.sdss.org/dr6/coverage/credits.html

\appendix

\section{Comparison with previous works}

The comparison of our results with previous works uses
a reduced model, because part of the data needed
for our full analysis is unpublished. Generally speaking,
previous works did not publish observed values of
background and total counts, $obsbkg$ and $obstot$, but 
just $obsn200=obstot-obsbkg/C$ and assumed the latter quantity
(an observed value) to be equal to $n200$ (the true value).

In the maxBCG work, such an identification leads to a significant 
bias, as discussed later.
In the other works, such identification is safer, but
authors systematically underestimate their uncertainties
either assuming that the mass-richness scaling has no intrinsic
scatter, or that the slope of the scaling is perfectly
known, for example when the intrinsic scatter is
derived. 

Rines et al. (2003) compute $N_{200}$ values for
a sample of 9 clusters with available 
caustic masses. Their scaling, derived by a
least-square fits, has {\it inverse} slope $0.70\pm0.09$. 
Our revised model now assumes that the observed richness
is Gaussian distributed with mean 
$n200$ and standard deviation $obserrn200$ and an
uniform prior on $n200$. In formulae, eq. 5, 6, 12 and 13 are
replaced by:

\begin{eqnarray}
obsn200_i & \sim & \mathcal{N}(n200_i, obserrn200^2_i) \\
n200_i & \sim & \mathcal{U}(0,\infty) 
\end{eqnarray}

With our model, which allows
an intrinsic scatter that their 
least-square fit does not, we found
(using their data kindly made available to us) a
slope of $1.23\pm 0.25$, in agreement
with our determination using a larger sample. The slope error
we found on their data is about twice
as large as that we found on our data (our sample is
6 times bigger) and is larger than their
quoted slope error, derived assuming no intrinsic 
scatter (and also no noisiness of mass errors). 
The 95\% confidence interval on the intrinsic scatter  
derived with this small sample 
largely depends on the adopted prior, in contrast to
what we find with our larger sample.

Muzzin et al. (2007) measure $N500$ for a sample of
15 clusters (one of which is discarded a-posteriori)
with dynamical masses (i.e.
coming from a velocity dispersion measurement). 
They use masses and richnesses
within $r_{500}$, $M500$ and
find a slope of $1.40\pm0.22$ with mass.
Their slope is at 
$1.6$ combined $\sigma$ from the slope we
derive for our sample. However,
their uncertainty on the found slope assumes
that no intrinsic scatter is there and
once one is allowed, their errors escalate
and the difference between the two slopes, in
terms of combined $\sigma$, decreases.

For 25 clusters in the Red Cluster Sequence Survey (RCS, Gladders 
et al. 2005), Blindert (2006)
computes the scaling between the RCS richness, $B_{gcR}$,
and velocity dispersion. Their richness only uses
red galaxies, as does ours.
We note however that their velocity dispersions $s$ are
derived from a small number of velocities ($<25$ for
about half the sample, vs our average of 208 velocities per
cluster) and thus have low 
reliability (Andreon et al. 2008; Andreon 2009; Gal et al. 2008). 
They found a slope of $0.75\pm0.57$, which is entirely
consistent with 
our, better determined, slope of $0.30\pm0.04$,
given Blindert (2006)'s large errors.

Johnston et al. (2007) stack maxBCG (Koester et al. 2007a) clusters, derive
masses from lensing and measure the scaling between richness and
the derived masses. For our present purposes, it is probably of
little relevance that their $obsn200$, 
counting by definition galaxies within $r_{200}$, counts instead
galaxies within $2 r_{200}$.
They obtain a slope of $1.28\pm0.04$, 
much more precise than our slope, which has an error
of $0.15$, at least at
first sight. Johnston et al.'s (2007) statistical analysis is quite
complex. Let us consider just a single
aspect, Johnston et al.
(2007) did not account for the difference between the observed value, 
$obsn200$ and the true value $n200$\footnote{This difference is
instead considered when the distribution of
$obsn200$ is used to constrain cosmological parameters in
Rozo et al. (2009a,b).}. Errors introduce a
scatter between $n200$ and $obsn200$ and,
because of the large abundance of clusters of low richness, 
the scatter brings many more low-richness clusters up
than high-richness clusters down. This implies that a given observed
richness, $obsn200$, many objects have
indeed a $n200<obsn200$. This selection effect,
usually called Malmquist or Eddington bias, is especially
large for the maxBCG clusters, whose observed richness is
as low as 3. Let us compute the Malmquist or Eddington
correction; in mathematical terms, 
$p(n200| obsn200)
\propto p( obsn200| n200) p( n200)$ 
(Bayes theorem). The cluster number
counts in Johnston et al. (2007) paper have a logarithmic slope of about 
$-3$ ($=\partial \log n /\partial \log n200$). This is the adopted
logarithmic slope of the
prior $p( n200)$. The likelihood, 
$p( obsn200| n200)$, is Poisson. Performing the algebra it
turns out that on average $n200 \sim obsn200-2$, as 
qualitatively expected.
If we now refit the Johnston et al. (2007) richness-mass data
using Malmquist corrected
values (i.e. using $obsn200-2$), 
we got a slope of $\approx 1.0$. This is about $7\sigma$ away from 
the quoted value, if we trust the slope error as published by 
Johnston et al. (2007), this shows that their slope is not robust
and their slope error is largely underestimated.
As stressed by Jeffreys (1938) and Eddington (1940), our
correction above has to be taken as an indication, 
by no means as a replacement of the correct analysis. It has
been presented only to give a glimpse about its size. 
Our finding that the
mass richness calibration by Johnston et al. (2007) is
more uncertain than claimed is also supported by the
result of a very similar but independent lensing analysis by
Mandelbaum et al. (2008a).

Lin et al. (2004) compute the richness-mass scaling
for a large cluster sample and found a slope about 1 $\sigma$ away from our
one, but with very small errors. However, their masses are 
derived from X-ray temperatures, in turn assumed to be perfectly known
(even in presence of large temperature errors), although they
have been
derived in heterogeneous ways (e.g. from measurements
performed in heterogeneous selected apertures, with or without
flagging the cool
core, etc.) from heterogeneous data/telescopes. Furthermore,
temperature-mass scaling is assumed to be perfectly
known, without any scatter and valid for clusters not 
in hydrostatic equilibrium, none of which is true. Similarly,
the radius within which richness is computed is estimated
from cluster temperature, assuming no scatter between
temperature and mass. Therefore, the 
small errors quoted by Lin et al. (2004) are found 
in an analysis where masses and radii are assumed perfectly known, in
spite of their significant noisiness and, possibly,
bias.

\section{Model listing}

In this section we give the listing of the full model. 

For the stochastic computation and for building the statistical
model we use Just Another Gibb Sampler 
(JAGS\footnote{http://calvin.iarc.fr/$\sim$martyn/software/jags/}, 
Plummer 2008). Eq 5 to 15
find an almost literal translation in JAGS,
Poisson, Normal and Uniform distributions become
{\texttt{dpois, dnorm, dunif}}, respectively. 
JAGS, following BUGS (Spiegelhalter et al. 1995), uses 
precisions, $prec = 1/\sigma^2$, in place of variances $\sigma^2$. 
Furthermore, it uses neperian logarithms, instead of decimal ones.
Eq. 6 has been rewritten using the property that the $\chi^2$
is a particular form of the Gamma distribution. 
Eq. 7 is split in two JAGS lines for a better reading.
The arrow symbol reads ``take the value of". 
{\texttt{obsvarlgM200}} is the square of $obserrlgM200$.

\begin{verbatim}
data
{
	nu <-6
}
model 
{
for (i in 1:length(obstot)) {
  obsbkg[i] ~ dpois(nbkg[i])
  obstot[i] ~ dpois(nbkg[i]/C[i]+n200[i]) 
  n200[i] ~ dunif(0,3000)
  nbkg[i] ~ dunif(0,3000)

  precy[i] ~ dgamma(1.0E-5,1.0E-5)
  obslgM200[i] ~ dnorm(lgM200[i],precy[i])
  obsvarlgM200[i] ~ dgamma(0.5*nu,0.5*nu*precy[i])

  z[i] <- alpha+14.5+beta*(log(n200[i])/2.30258-1.5)
  lgM200[i] ~ dnorm(z[i], prec.intrscat)
  }
intrscat <- 1/sqrt(prec.intrscat)
prec.intrscat ~ dgamma(1.0E-5,1.0E-5)
alpha ~ dnorm(0.0,1.0E-4)
beta ~ dt(0,1,1)
}
  
\end{verbatim}

In order to evaluate eq. 4, i.e. to determine
the uncertainty of the predicted mass,
we simply need to add to the data
file the list of clusters for which we want predictions. 
In this paper we used the same sample, as mentioned in Sec 6,
which is, as a result, listed twice in the data file, the
second time with  ``NA'' (``not available'') values of mass indicating
to the program that they should be estimated.

\bsp

\label{lastpage}

\end{document}